\documentstyle[12pt]{article}
\begin{document}
\begin{center}

{\large\bf Wigner Functions for harmonic oscillator in
noncommutative phase space} \vskip 1cm Jianhua Wang$^{a,d}$, Kang
Li$^{b,d}$ and Sayipjamal Dulat$^{c,d}$
\\\vskip 1cm

$^a$ Department of Physics, Shaanxi University  of
Technology, Hanzhong, 723001, China\\
$^b $Department of Physics, Hangzhou Normal University, Hangzhou,
310036, China\\
$^c$School of Physics Science and Technology, Xinjiang University,
Urumqi, 830046, China\\
$^d$Kavli Institute for Theoretical Physics China , CAS, Beijing
100190, China
 \vskip 0.5cm
\end{center}

\begin{abstract}
\noindent We study the Wigner Function  in non-commutative quantum
mechanics. By solving the time independent Schr\"{o}dinger equation
both on a non-commutative (NC) space  and a non-commutative phase
space, we obtain the Wigner Function for the harmonic oscillator  on
NC space and NC phase space respectively.

 PACS number(s): 02.40.Gh, 03.65.Ca,
03.65.-w
\end{abstract}

\section{Introduction}
The study of physics effects on a NC space and a NC phase space has
attracted much attention in recent years. Because the effects of the
space non-commutativity may become significant in the string scale.
Besides the field theory, there are many papers devoted to the study
of various aspects of quantum mechanics on a NC space and a NC phase
space with usual (commutative) time coordinate
\cite{AB-2}-\cite{kang}. For example, the Aharonov-Bohm phase on a
NC space and a NC phase space has been studied in Refs.
\cite{AB-2}-\cite{AB-4}. The Aharonov-Casher phase for a spin-$1/2$
and spin-1 particle on a NC space and a NC phase space has been
studied in Refs.\cite{AC-NCS spin-half}-\cite{AC-NCS spin-one}.
Landau problem and HMW effect both on  NC space and   NC phase space
were studied in Refs.\cite{NC-Landau}\cite{NC-HMW}. There were some
studies concerning the quantum Hall effect on  NC space
\cite{NC-QH1} and   NC phase space\cite{NC-QH2}. Ref.\cite{jing}
studied Wigner function   for the non-Hamiltonian systems on a NC
space. Wigner function is a very important function not only because
it is equivalent to the Schr$\ddot{o}$dinger wave function in
quantum mechanics, but also it relates to the quantum observation,
so the further study of NC wigner function is very important and
useful. In this paper we study the effect of the noncommutativity
via Wigner function for the harmonic oscillator. The article is
organized as follows: In section \ref{WF}, we review the Wigner
distribution function, as an example we calculate Wigner Function
for two dimensional Harmonic oscillator. In section \ref{WF-NCS}, we
study the Wigner functions for the Harmonic oscillator in NC space.
In section \ref{WF-NCPS}, by using a generalized Bopp's shift, we
deduce Wigner functions for the Harmonic oscillator on NC phase
space. Conclusions are given in the last section.

\section{Wigner functions and Harmonic oscillator}\label{WF}
There are three types of formulation quantum mechanics. Namely,
standard operator quantization, which was developed by Schrodinger,
Dirac, and Heisenberg.The second is the path integral quantization,
which was constructed by Feynman. The last one is the phase space
formulation of quantum mechanics (also known as the Moyal
quantization or deformation quantization), which was due to
Wigner\cite{wigner}, which is less well known, but which is useful
in many areas of physics. For example, it is useful in describing
quantum transport process in phase space, and has importance in
quantum optics, nuclear physics, condensed matter, M-theory,
noncommutative geometry, and matrix models.  There are no operators
in this formulation of quantum mechanics. Observables and transition
amplitudes are phase space integrals of classical number functions,
which compose via the star product, and they weighted by the Wigner
function , as in statistical mechanics. Wigner constructed a
distribution function, which is real, but not everywhere positive,
from the quantum-mechanical wave function. Moyal then gave the
evolution equation for this distribution, introducing his famous
bracket\cite{moyal}. The definition of Wigner probability function
of the simultaneous values of ${\bf x}$ for the coordinates and
${\bf p}$ for the momenta in 2d- dimensional phase space in terms of
the wave function $\psi({\bf x})$ of Schr$\ddot{o}$dinger equation
$\hat H(\hat{\bf x},\hat{\bf p}) \psi({\bf x}) = E\psi({\bf x})$,
($\hat H$ is the Hamiltonian operator, where the coordinates
$\hat{\bf x}$, and  momenta $\hat{\bf p}$ satisfy standard
commutation relation: $[\hat{x}_{i},\hat{p}_{j}]=i\hbar
\delta_{ij}$) is
\begin{equation}\label{eq-1}
W({\bf x},{\bf p})=\frac{1}{(2\pi)^d} \int_{-\infty}^{+\infty}
 d{\bf y} e^{-i{\bf y.p}}\psi^*({\bf x}-{\hbar\over 2} {\bf y}) \psi({\bf x}+{\hbar\over 2} {\bf y}),
\end{equation}
where $\psi^*({\bf x})$ stands for complex conjugate of $\psi({\bf
x})$.
 One can also obtain time-independent pure state Wigner function by solving directly star-genvalvue equation \cite{curtright}
\begin{equation}\label{eq-2}
H({\bf x},{\bf p}) \ast_{\hbar} W ({\bf x},{\bf p})
 =W({\bf x},{\bf p})\ast_{\hbar} H({\bf x},{\bf p}) =EW({\bf x},{\bf p}),
\end{equation}
where the associative star-product is
\begin{equation}\label{eq-3}
\ast_{\hbar} \equiv e^{{i\hbar \over 2}
(\overleftarrow{\partial}_{\bf x}\cdot
\overrightarrow{\partial}_{\bf p} -\overleftarrow{\partial}_{\bf
p}\cdot \overrightarrow{\partial}_{\bf x} )};
\end{equation}
and $H({\bf x},{\bf p})$ is the classical Hamiltonian function
corresponding to $\hat H$. The star-product encodes the entire
quantum mechanical action. Recalling the action of a translation
operator, the star-product induces "Bopp" shifts
\begin{eqnarray}\label{eq-4}
f({\bf x},{\bf p}) \ast_{\hbar} g({\bf x},{\bf p}) &=& f\left( {\bf
x}+{i\hbar\over 2}\overrightarrow{\partial}_{\bf p},{\bf p} -
{i\hbar\over 2}\overrightarrow{\partial}_{\bf x} \right ) g({\bf
x},{\bf p})\nonumber\\
&=& f\left({\bf x}, {\bf p}-{i\hbar\over
2}\overrightarrow{\partial}_{\bf x}\right) g\left({\bf x}, {\bf p}
+{i\hbar\over 2}\overleftarrow{\partial}_{\bf x}\right).
\end{eqnarray}
Note that $f({\bf x},{\bf p}) \ast_{\hbar} g({\bf x},{\bf p})$
denotes quantum deformation of a usual commutative product of
functions $f\cdot g$. In conclusion there are two equivalent ways to
get Wigner functions, namely equation (\ref{eq-1}) or (\ref{eq-2}).

To illustrate the approach of equation (\ref{eq-2}), we look at the
2-dimensional harmonic oscillator described by the following
Hamiltonian (with $m=1,\omega=1$)
\begin{equation}\label{eq2-2}
  H(x,p) = \frac{1}{2} [( p_1^2 + x_1^2)+( p_2^2 + x_2^2)] \ .
\end{equation}
Now, let us study the corresponding eigenvalue problem of
(\ref{eq-2}). Equation $H \ast W = EW$ gives
\begin{equation}\label{eq2-3}
  [ x_1^2 + p_1^2 - \frac{\hbar^2}{4} ( \partial^2_{x_1} +
  \partial^2_{p_1}) + x_2^2 + p_2^2 - \frac{\hbar^2}{4} ( \partial^2_{x_2} +
  \partial^2_{p_2}) +\frac{i\hbar}{2} (x_1\partial_{p_1} - p_1\partial_{x_1}+x_2\partial_{p_2} - p_2\partial_{x_2})-
  2E ] W = 0\  ,
\end{equation}
whereas $W \ast H = EW$
\begin{equation}\label{eq2-4}
  [ x_1^2 + p_1^2 - \frac{\hbar^2}{4} ( \partial^2_{x_1} +
  \partial^2_{p_1}) - p_1\partial_{x_1}) + x_2^2 + p_2^2 - \frac{\hbar^2}{4} ( \partial^2_{x_2} +
  \partial^2_{p_2}) -\frac{i\hbar}{2} (x_1\partial_{p_1} - p_1\partial_{x_1}+x_2\partial_{p_2} - p_2\partial_{x_2}) -
  2E ] W = 0\ .
\end{equation}
Therefore
\begin{equation}\label{eq2-5}
(x_1\partial_{p_1} - p_1\partial_{x_1}+x_2\partial_{p_2} -
p_2\partial_{x_2})\, W = 0 \ ,
\end{equation}
which means that $W$ is a zero-mode of the Koopman operator
$L_{H_o}$ \cite{dariusz}. Taking into account eq.(\ref{eq2-3}) and
eq.(\ref{eq2-4})
 we obtain
\begin{equation}\label{eq2-6}
  [ x_1^2 + p_1^2 - \frac{\hbar^2}{4} ( \partial^2_{x_1} +
  \partial^2_{p_1}) + x_2^2 + p_2^2 - \frac{\hbar^2}{4} ( \partial^2_{x_2} +
  \partial^2_{p_2}) -
  2E ]  W = 0\ ,
\end{equation}
Introducing two new variables $\xi$ and $\eta$
\begin{equation}\label{eq2-7}
  \xi := \frac{2}{\hbar}\, (x_1^2 + p_1^2),\ \ \ \eta := \frac{2}{\hbar}\, (x_2^2 +
  p_2^2),
\end{equation}
equation (\ref{eq2-6}) may be rewritten as follows
\begin{equation}\label{eq2-8}
  \left[ \frac{\xi}{4} - \xi\partial^2_\xi - \partial_\xi
   + \frac{\eta}{4} - \eta\partial^2_\eta - \partial_\eta \right] W(\xi,\eta) = E  W(\xi,\eta),
\end{equation}
Let $W(\xi,\eta)= W(\xi) W(\eta)$, $E=E_1+E_2$, we have
\begin{equation}\label{eq2-9}
\left[ \frac{\xi}{4} - \xi\partial^2_\xi - \partial_\xi - E_1\right]
W(\xi) = 0\ ,
\end{equation}
and
\begin{equation}\label{eq2-10}
\left[ \frac{\eta}{4} - \eta\partial^2_\eta - \partial_\eta -
E_2\right] W(\eta) = 0\ .
\end{equation}
By defining $W(\xi)$ as
\begin{equation}\label{eq2-11}
  W(\xi) =: e^{-\xi/2}\, L(\xi)\ ,
\end{equation}
we rewrite equation (\ref{eq2-9}) as
\begin{equation}\label{eq2-12}
  \left[   \xi \partial^2_{\xi} + (1-\xi)\partial_\xi +
  \frac{E}{\hbar} - \frac 12 \right] L(\xi) = 0 \ ,
\end{equation}
solutions of the above  equation are the Laguerre's polynomials
\begin{equation}\label{eq2-13}
  L_m(\xi) = \frac{1}{m!} \, e^\xi \partial_\xi(e^{-\xi}\, \xi^m)\ ,
\end{equation}
for $m= E_1/\hbar - 1/2=0,1,\ldots\,$. The corresponding Wigner
functions $W_m$ are
\begin{equation}\label{eq2-14}
  W_m = \frac{(-1)^m}{\pi \hbar} \, e^{-\xi/2} \, L_m(\xi) \ .
\end{equation}
Similarly, for $n= E_2/\hbar - 1/2=0,1,\ldots\,$. The corresponding
Wigner functions $W_n$ are
\begin{equation}\label{eq2-15}
  W_n= \frac{(-1)^n}{\pi \hbar} \, e^{-\eta/2} \, L_n(\eta) \ .
\end{equation}
Thus we have
\begin{equation}\label{eq2-16}
  W_{mn} = \frac{(-1)^{m+n}}{(\pi \hbar)^2} \, e^{{-(\xi+\eta)}/2} \, L_m(\xi)L_n(\eta) \ .
\end{equation}
Substituting equation (\ref{eq2-7}) into equation (\ref{eq2-16}), we
may rewrite the Wigner functions $W_{mn}$ for 2-dimensional harmonic
oscillator as follows
\begin{equation}\label{eq2-17}
  W_{mn}(x_1,p_1,x_2,p_2) = \frac{(-1)^{m+n}}{(\pi \hbar)^2} \, e^{{-(x_1^2 + p_1^2+x_2^2 +p_2^2)}/{\hbar}} \,
   L_m[\frac{2}{\hbar}\, (x_1^2 + p_1^2)]L_n[\frac{2}{\hbar}\, (x_2^2 +p_2^2)] \ .
\end{equation}
When $n=0$, $m=0$, we have
\begin{equation}\label{eq2-18}
  W_{00}  = \frac{1}{(\pi \hbar)^2} \,
  e^{-(x_1^2 + p_1^2+x_2^2 +p_2^2)/\hbar}\ ,
\end{equation}
The Wigner distribution function is non-negative  Gaussian
distribution function.  However, in the classical limit $\hbar
\longrightarrow 0$ all  Wigner
 functions $W_{mn}$ tend to well defined classical probability
 distributions. For example
\begin{equation}\label{eq2-19}
  W_{00}(x_1,p_1,x_2,p_2) \ \longrightarrow \ \delta(x_1)\delta(p_1) \delta(x_2)\delta(p_2)\ .
\end{equation}

\section{Wigner functions for Harmonic oscillator on NC space}\label{WF-NCS}

On a NC plane coordinates $\hat{x}^{nc}_i$  and momenta
$\hat{p}^{nc}_i$  $(i=1,2)$ operators satisfy the following
commutation relations
\begin{equation}\label{eq3-1}
~[\hat{x}^{nc}_{i},\hat{x}^{nc}_{j}]=i\theta_{ij},~~~
[\hat{p}_{i},\hat{p}_{j}]=0,~~~[\hat{x}^{nc}_{i},\hat{p}_{j}]=i\hbar
\delta_{ij}.
\end{equation}
The Schr$\ddot{o}$dinger equation on a NC space is
\begin{equation}\label{eq3-2}
 \hat{ H}(\hat{\bf x},\hat{\bf p}
)\ast_\theta \psi^{nc}({\bf x})=E \;\psi^{nc} ({\bf x}),
\end{equation}
where the Moyal-Weyl (or star) product  is defined as
\begin{equation}\label{eq3-3}
 \ast_\theta   = e^{\frac{i \theta }{2}
 (\overleftarrow{\partial}_{x_1} \overrightarrow{\partial}_{x_2} - \overleftarrow{\partial}_{x_2} \overrightarrow{\partial}_{x_1})},
\end{equation}
After obtaining $\psi^{nc}({\bf x})$ from (\ref{eq3-2}), the Wigner
function on a NC space is
\begin{equation}\label{eq3-4}
W^{nc}({\bf x},{\bf p})={1\over (2\pi)^2}\int\! d{\bf y}~e^{-i{\bf
y}\cdot {\bf p}}~\psi^{nc} ({\bf x}+{\hbar\over2} {\bf y} )
\ast_\theta \psi^{\ast nc}({\bf x}-{\hbar\over2} {\bf y}) ,
\end{equation}
Alternatively one can also get the NC space Wigner function by
solving the following star-genvalue equation
\begin{equation}\label{eq3-5}
H({\bf x},{\bf p}) \ast W^{nc}({\bf x},{\bf p})
 =W^{nc}({\bf x},{\bf p})\ast H({\bf x},{\bf p}) =E\;W^{nc}({\bf x},{\bf p}),
\end{equation}
where
\begin{equation}\label{eq3-stars}
\ast=\ast_{\hbar}~\ast_\theta =exp\left({ \frac{i \hbar
}{2}\sum_{i=1}^{2}(\overleftarrow{\partial}_{x_i}
\overrightarrow{\partial}_{p_i} - \overleftarrow{\partial}_{p_i}
\overrightarrow{\partial}_{x_i}) + \frac{i \theta }{2}
 (\overleftarrow{\partial}_{x_1} \overrightarrow{\partial}_{x_2} - \overleftarrow{\partial}_{x_2}
 \overrightarrow{\partial}_{x_1})}\right)
\end{equation}
Instead of solving the NC space Schr$\ddot{o}$dinger equation by
using the star product procedure, we use a Bopp's shift method, that
is, we replace the $\ast_\theta$-product in Schr$\ddot{o}$dinger
equation with usual product by making  a Bopp's shift
\begin{equation}\label{eq3-6}
 \hat{x}^{nc}_{i}=  \hat{x}_{i}-\frac{1}{2\hbar}\theta_{ij} \hat{p}_{j} ,
 ~~\hat{p}^{nc}_i=\hat{p}_i, \hspace{1cm} i = 1, 2.
\end{equation}
where $\theta_{ij}= \theta\epsilon_{ij}$. Then the equation
(\ref{eq3-2}) takes the following form
\begin{equation}\label{eq3-7}
 \hat{ H}(\hat{{\bf x}}^{nc},\hat{{\bf p}}^{nc})\psi^{nc}( {\bf x})=E\;\psi^{nc}({\bf x}),
\end{equation}
Such that equation (\ref{eq3-5}) can be rewritten as
\begin{equation}\label{eq3-8}
H({\bf x}^{nc},{\bf p}^{nc}) \ast_{\hbar} W^{nc}({\bf x},{\bf p})
 =W^{nc}({\bf x},{\bf p})\ast_{\hbar} H({\bf x}^{nc},{\bf p}^{nc}) =E\; W^{nc}({\bf x},{\bf p}),
\end{equation}
where
\begin{equation}\label{eq3-9}
 x^{nc}_i=  x_{i}-\frac{1}{2\hbar}\theta_{ij}p_{j} ,
 ~~ p^{nc}_i=p_i, \hspace{1cm} i = 1, 2.
\end{equation}
By comparing equation (\ref{eq-2}) with (\ref{eq3-8}) we obtain
\begin{equation}\label{eq3-10}
 W^{nc}_{mn}(x_1,p_1,x_2,p_2) =W(x_1\rightarrow x^{nc}_1, p_1\rightarrow p^{nc}_1,x_2\rightarrow x^{nc}_2, p_2\rightarrow
 p^{nc}_2)
\end{equation}
Therefore
\begin{eqnarray}\label{eq3-11}
  W^{nc}_{mn}(x_1,p_1,x_2,p_2) &=& \frac{(-1)^{m+n}}{(\pi \hbar)^2} \, e^{{-((x^{nc}_1)^2 + (p^{nc}_1)^2+ (x^{nc}_2)^2 + (p^{nc}_2)^2)}/{\hbar}}
  \nonumber\\
   &&L_m[\frac{2}{\hbar}\, ((x^{nc}_1)^2 + (p^{nc}_1)^2)]L_n[\frac{2}{\hbar}\, ((x^{nc}_2)^2 + (p^{nc}_2)^2))] \ .
\end{eqnarray}
Inserting (\ref{eq3-9}) into (\ref{eq3-11}), and neglecting term
with $\theta^2$, we have
\begin{eqnarray}\label{eq3-12}
W^{nc}_{mn}(x_1,p_1,x_2,p_2) &=& \frac{(-1)^{m+n}}{(\pi \hbar)^2}
e^{{-[(x_1^2 +p_1^2+x_2^2 +p_2^2)- \frac{\theta}{\hbar}(x_1 p_2-x_2
p_1)
 ]}/{\hbar}}\nonumber\\
&& L_m[\frac{2}{\hbar}\,( x_1^2 +
p_1^2-\frac{\theta}{\hbar}x_1p_2)]L_n[\frac{2}{\hbar}\,( x_2^2 +
p_2^2-\frac{\theta}{\hbar}x_2p_1)],
\end{eqnarray}
this is the Wigner functions for 2-dimensional Harmonic oscillator
in NC space. For $n=0$, $m=0$, one has
\begin{equation}\label{eq3-13}
W^{nc}_{00}(x_1,p_1,x_2,p_2) = \frac{1}{(\pi \hbar)^2} e^{{-[(x_1^2
+p_1^2+x_2^2 +p_2^2)- \frac{\theta}{\hbar}(x_1 p_2-x_2 p_1)
 ]}/{\hbar}}
 \end{equation}

\section{Wigner functions for Harmonic oscillator in NC phase space}\label{WF-NCPS}

The case of both space-space and momentum-momentum noncommuting
\cite{kochan}\cite{kang} is different from the case of only
space-space noncommuting.  Thus on a NC phsae space, not only the
coordinate operators are noncommutative as in (\ref{eq3-1}), but
also the momentum operators in equation (\ref{eq3-1}) satisfy the
following commutation relations
\begin{equation}\label{eq4-1}
[\hat{p}^{nc}_{i},\hat{p}^{nc}_{j}]=i\bar{\theta}_{ij},\hspace{1cm}
i,j = 1,2.
\end{equation}
Here $\{\bar{\theta}_{ij}\}$  is a totally antisymmetric matrices
which represent the noncommutative property among the momenta on a
NC phase space, and play analogous role to $\hbar$ in the usual
quantum mechanics. The Schr$\ddot{o}$dinger  equation on a NC phase
space is written as
\begin{equation}\label{eq4-2}
 \hat{H}(\hat{{\bf x }},\hat{{\bf p}}) \ast_{\theta} \ast_{\bar{\theta}} \; \psi^{ncps}( {\bf x})=E \;\psi^{ncps}({\bf x}),
\end{equation}
where the $\ast_ {\bar{\theta}}$-product in Eq.(\ref{eq4-2}), for NC
phase space, is defined by
\begin{equation}\label{eq4-3}
\ast_ {\bar{\theta}} = e^{ \frac{i\bar{\theta}}{2}
 (\overleftarrow{\partial}_{p_1} \overrightarrow{\partial}_{p_2} - \overleftarrow{\partial}_{p_2}
 \overrightarrow{\partial}_{p_1})}.
\end{equation}
Wigner function on a NC phase space is written as
\begin{equation}\label{eq4-4}
W^{ncps}({\bf x},{\bf p})={1\over (2\pi)^2}\int\! d{\bf y}~e^{-i{\bf
y}\cdot{\bf p}}~\psi^{ncps} ({\bf x}+{\hbar\over2}{\bf y} )
\ast_{\theta} \ast_{\bar{\theta}} \psi^{\ast ncps}({\bf
x}-{\hbar\over2}{\bf y}) ,
\end{equation}
The $\ast$-genvalue equation on a NC phase space is
\begin{equation}\label{eq4-5}
H({\bf x},{\bf p}) \ast W({\bf x},{\bf p})
 =W({\bf x},{\bf p})\ast H({\bf x},{\bf p}) = E\;W({\bf x},{\bf p}),
\end{equation}
where $\ast=\ast_{\hbar}\ast_{\theta} \ast_{\bar{\theta}}$.

Instead of solving the NC phase space Schr$\ddot{o}$dinger equation,
we use a Bopp's shift method, that is, we replace the $\ast_\theta
\ast_{\bar{\theta}}$-product in Schr$\ddot{o}$dinger equation with
usual product by making the following a Bopp's shift
\begin{eqnarray}\label{eq4-6}
 {\hat x}^{ncps}_i &=&   \alpha\hat{x}_{i} - \frac{1}{2 \alpha\hbar}\theta_{ij} \hat{p}_j,\nonumber\\
 {\hat p}^{ncps}_i &=&
\alpha\hat{p}_i+\frac{1}{2\alpha\hbar}\bar{\theta}_{ij}\hat{x}_j,
\hspace{1cm} i = 1, 2.
\end{eqnarray}
where $\bar{\theta}_{ij}= \bar{\theta}\epsilon_{ij}$,
$\theta\bar\theta = 4\hbar^2\alpha^2(1-\alpha^2)$, $\alpha =
1-\frac{\theta\bar\theta}{8\hbar^2}= 1+ O(\theta^2)$. Hereafter we
choose $\alpha = 1$. Then the equation (\ref{eq4-2}) takes the
following form
\begin{equation}\label{eq4-7}
 \hat{ H}(\hat{{\bf x}}^{ncps},\hat{{\bf p}}^{ncps})\psi^{ncps}( {\bf x})=E\;\psi^{ncps}({\bf x}),
\end{equation}
Such that equation (\ref{eq4-5}) can be rewritten as
\begin{equation}\label{eq4-8}
H({\bf x}^{ncps},{\bf p}^{ncps}) \ast_{\hbar} W^{ncps}({\bf x},{\bf
p})
 =W^{ncps}({\bf x},{\bf p})\ast_{\hbar} H({\bf x}^{ncps},{\bf p}^{ncps}) =E\; W^{ncps}({\bf x},{\bf p}),
\end{equation}
where
\begin{eqnarray}\label{eq4-9}
 x^{ncps}_i &=&  x_{i}-\frac{1}{2\hbar}\theta_{ij}p_{j} ,\nonumber\\
 ~~ p^{ncps}_i &=& p_i + \frac{1}{2\hbar}\bar{\theta}_{ij}x_j, \hspace{1cm} i
= 1, 2.
\end{eqnarray}
By comparing equation (\ref{eq-2}) with (\ref{eq4-8}) we obtain
\begin{equation}\label{eq4-10}
 W^{ncps}_{mn}(x_1,p_1,x_2,p_2) =W(x_1\rightarrow x^{ncps}_1, p_1\rightarrow p^{ncps}_1,x_2\rightarrow x^{ncps}_2, p_2\rightarrow
 p^{ncps}_2)
\end{equation}
Therefore
\begin{eqnarray}\label{eq4-11}
  W^{ncps}_{mn}(x_1,p_1,x_2,p_2) &=& \frac{(-1)^{m+n}}{(\pi \hbar)^2} \, e^{{-((x^{ncps}_1)^2 + (p^{ncps}_1)^2+ (x^{ncps}_2)^2 + (p^{ncps}_2)^2)}/{\hbar}}
  \nonumber\\
   &&L_m[\frac{2}{\hbar}\, ((x^{ncps}_1)^2 + (p^{ncps}_1)^2)]L_n[\frac{2}{\hbar}\, ((x^{ncps}_2)^2 + (p^{ncps}_2)^2))] \ .
\end{eqnarray}
Inserting Eq.(\ref{eq4-9}) into Eq.(\ref{eq4-11}), and neglecting
term with $\theta^2$, and $\bar{\theta}^2$,  we have
\begin{eqnarray}\label{eq4-12}
W^{ncps}_{mn}(x_1,p_1,x_2,p_2) &=& \frac{(-1)^{m+n}}{(\pi \hbar)^2}
e^{{-[x_1^2 +p_1^2+ x_2^2 +p_2^2 -
\frac{\theta+\bar{\theta}}{\hbar}( x_1 p_2-  x_2 p_1)
  ]}/{\hbar}}\nonumber\\
&& L_m\{\frac{2}{\hbar}[  x_1^2 + p_1^2-\frac{1}{\hbar}(\theta
x_1p_2-\bar{\theta} x_2p_1)]\}\nonumber\\
&& L_n\{\frac{2}{\hbar}[ x_2^2 + p_2^2-\frac{1}{\hbar}(\bar{\theta
}x_1p_2-\theta x_2p_1)]\} \ .
\end{eqnarray}
this is the Wigner functions for the Harmonic oscillator on a NC
phase space. Again Wigner functions which corresponds to the ground
sate wave function is given by
\begin{equation}\label{eq4-13}
W^{ncps}_{00}(x_1,p_1,x_2,p_2) = \frac{1}{(\pi \hbar)^2} e^{{-[x_1^2
+p_1^2+x_2^2 +p_2^2- \frac{\theta+\bar{\theta}}{\hbar}( x_1 p_2- x_2
p_1)
  ]}/{\hbar}}.
 \end{equation}

\section{Conclusion remarks }

In this paper, we study the Wigner functions for the Harmonic
oscillator
 both on a noncommutative space and a noncommutative phase space. Instead
of doing tedious star product calculation, we use the "shift"
method, i.e. the star product in equations (\ref{eq3-2}) and
(\ref{eq4-2}) can be replaced by Bopp's shift equations
  (\ref{eq3-7}) for NC space and (\ref{eq4-7}) for NC
phase space. These shifts are
 equal to the star product. The additional terms in (\ref{eq3-12}) on a
 NC space and in (\ref{eq4-12}) on a NC phase space are related to the
non-commutativity of space and phase space. This effect is expected
to be tested at a very high energy level, and the experimental
observation of the effect remains to be further studied.

The method we use in this paper may also be employed to other
physics problem on NC space and NC phase space. The further study on
the issue will be reported in our forthcoming papers.

\section{Acknowledgments}  We
would like to thank Prof. Yue-Liang Wu for his kind invitation and
warm hospitality during our visit at the KITPC. This work is
supported by the National Natural Science Foundation of China (
10875035 as well as 10665001) and the Natural Science Foundation of
Zhejiang Provence (Y607437).

\end{document}